\documentclass[acmlarge]{acmart}
\usepackage{hyperref}
\usepackage{hyperxmp}
\usepackage{booktabs} 
\usepackage{siunitx} 
\AtBeginDocument{%
  \providecommand\BibTeX{{%
    \normalfont B\kern-0.5em{\scshape i\kern-0.25em b}\kern-0.8em\TeX}}}

\setcopyright{acmcopyright}
\copyrightyear{2018}
\acmYear{2018}
\acmDOI{XXXXXXX.XXXXXXX}

\acmConference[Proc. ACM Interact. Mob. Wearable Ubiquitous Technol."]{Make sure to enter the correct
  conference title from your rights confirmation emai}{volume 0 issue 0}{Woodstock, NY}
\acmPrice{15.00}
\acmISBN{978-1-4503-XXXX-X/18/06}

\usepackage[normalem]{ulem}

\def\markup{1}

\if\markup1

\else

\newcommand{\st}[1]{}
\newcommand{\sout}[1]{}
\fi
\begin{document}

\title[SilverCycling]{SilverCycling: Exploring the Impact of Bike-Based Locomotion on Spatial Orientation for Older Adults in VR}


\author{Qiongyan CHEN}
\affiliation{
  \institution{Computational Media and Arts Thrust}
  \institution{The Hong Kong University of Science and Technology (Guangzhou)}
  \city{Guangzhou}
  \country{China}
  }
\email{qchen580@connnect.hkust-gz.edu.cn}

\author{Zhiqing WU}
\affiliation{
  \institution{Computational Media and Arts Thrust}
  \institution{The Hong Kong University of Science and Technology (Guangzhou)}
  \city{Guangzhou}
  \country{China}
  }
\email{zwu755@connect.hkust-gz.edu.cn}

\author{Yucheng LIU}
\affiliation{
  \institution{Computational Media and Arts Thrust}
  \institution{The Hong Kong University of Science and Technology (Guangzhou)}
  \city{Guangzhou}
  \country{China}
  }
\email{yliu428@connect.hkust-gz.edu.cn}

\author{Lei Han}
\affiliation{
  \institution{Computational Media and Arts Thrust}
  \institution{The Hong Kong University of Science and Technology (Guangzhou)}
  \city{Guangzhou}
  \country{China}
  }
\email{lhan229@connect.hkust-gz.edu.cn}

\author{Zisu LI}
\affiliation{
  \institution{IIP (Computational Media and Arts)}
  \institution{The Hong Kong University of Science and Technology}
  \city{Hong Kong}
  \country{China}
  }
\email{zlihe@connect.ust.hk}

\author{Ge Lin KAN}
\affiliation{
  \institution{Urban Governance and Design Thrust}
  \institution{The Hong Kong University of Science and Technology (Guangzhou)}
  \city{Guangzhou}
  \country{China}
  }
\email{gelin@hkust-gz.edu.cn}

\author{Mingming Fan}
\authornote{Corresponding Author}
\affiliation{
  \institution{Computational Media and Arts}
  \institution{The Hong Kong University of Science and Technology (Guangzhou)}
  \city{Guangzhou}
  \country{China}
}
\affiliation{
  \institution{Integrative Systems and Design \& Computer Science and Engineering}
  \institution{The Hong Kong University of Science and Technology}
  \city{Hong Kong SAR}
  \country{China}
}
\email{mingmingfan@ust.hk}

\renewcommand{\shortauthors}{Qiongyan, et al.}

\begin{abstract}
Spatial orientation is essential for people to effectively navigate and interact with the environment in everyday life. With age-related cognitive decline, providing VR locomotion techniques with better spatial orientation performance for older adults becomes important. Such advancements not only make VR more accessible to older adults but also enable them to reap the potential health benefits of VR technology. Natural motion-based locomotion has been shown its effective in enhancing younger users' performance in VR navigation tasks that require spatial orientation. However, there is a lack of understanding regarding the impact of natural motion-based locomotion on spatial orientation for older adults in VR. To address this gap, we selected the SilverCycling system, a VR bike-based locomotion technique that we developed, as a representative of natural motion-based locomotion, guided by findings from our pilot study. We conducted a user study with 16 older adults to compare SilverCycling with the joystick-based controller. The findings suggest SilverCycling potential to significantly enhance spatial orientation in the open-road urban environment for older adults, offering a better user experience. Based on our findings, we identify key factors influencing spatial orientation and propose design recommendations to make VR locomotion more accessible and user-friendly for older adults.

\end{abstract}

\ccsdesc[500]{Computer systems organization~Embedded systems}
\ccsdesc[300]{Computer systems organization~Redundancy}
\ccsdesc{Computer systems organization~Robotics}
\ccsdesc[100]{Networks~Network reliability}

\keywords{Virtual/Augmented Reality, Older Adults, Locomotion, Bike, Spatial Orientation}


\received{20 February 2007}
\received[revised]{12 March 2009}
\received[accepted]{5 June 2009}


\maketitle
\section{INTRODUCTION}
Older adults tend to have reduced spatial orientation and weaker muscle strength than younger adults \cite{devlin2001mind, moffat2006age}. Designing locomotion studies for older adults, therefore should tailor their needs, addressing concerns such as enhancing spatial orientation, ensuring safety, and accommodating diminished muscle strength. However, most research of different VR locomotion techniques on spatial orientation targets younger adults \cite{bowman1998methodology, Langbehn2018, kim2021evaluation, freiwald2020walking}. Little is known about how these techniques impact older adults’ spatial orientation and overall experience in VR. This study aims to bridge this gap by assessing the effectiveness of natural motion-based locomotion methods for enhancing the VR experience of older adults.

Natural locomotion methods have been shown to enhance spatial orientation more effectively than non-natural techniques, as they provide additional motion cues, such as proprioceptive feedback and vestibular cues \cite{chance1998locomotion, wilson2016vr}. Among these, real walking stands out as the most natural method that significantly enhance spatial performance in VR navigation \cite{wilson2016vr}. However, the practical implementation of real walking in VR is hindered by high costs and space constraints \cite{wilson2016vr}. Thus, various motion-based locomotion techniques have been developed to mitigate these issues while improving spatial orientation and user experience \cite{boletsis2017new, di2021locomotion}. However, motion-based locomotion techniques with different interaction fidelity can impose varying cognitive and physical loads, impacting users' spatial orientation and overall experiences \cite{bowman1998methodology, nabiyouni2015comparing}. This makes it essential to evaluate the performance of motion-based locomotion techniques, especially for older adults.

Walking-based locomotion techniques, such as Redirected Walking (RDW), walking-in-place (WIP), and omnidirectional walking (ODW) could potentially enhance navigation task performance in VR environments, may not be appropriate for older adults \cite{Langbehn2018, wilson2016vr, ruddle2009benefits}. For instance, RDW closely mimics natural walking could enhance spatial learning and reduce navigation errors, but it limits the user’s freedom in perceiving space, thus posing safety risks for older adults \cite{ruddle2009benefits, kontio2023feel}. Compared to walking-based locomotion, we opted for bike-based locomotion because it offers a safer, seated experience that is particularly suitable for older adults. Additionally, cycling is a familiar activity among contemporary older adults in many countries \cite{oosterhuis2016cycling, winters2015grew}. Moreover, studies have shown that cycling can effectively simulate walking movements, yielding good spatial performance in VR \cite{freiwald2020walking}. 

Despite the increasing focus on how different VR locomotion techniques affect spatial orientation, most research targets younger adults \cite{bowman1998methodology, Langbehn2018, kim2021evaluation, freiwald2020walking}. Little is known about how these techniques impact older adults' spatial orientation and overall experience in VR. This study aims to bridge this gap by assessing the effectiveness of natural motion-based locomotion methods for enhancing the VR experience of older adults.

In this work, we took the first step towards exploring this problem. Specifically, we sought to answer the research question (RQ): \textbf{How does natural motion-based locomotion, such as bike-based locomotion, affect spatial orientation and interaction experience among older adults?}

To address our RQ, we selected representative natural motion-based locomotion methods, including bike-based locomotion and walking-based locomotion (WIP and ODW), to further investigate their performance. For bike-based locomotion, we developed the SilverCycling system, which provided intuitive turning and forward motion functionalities tailored to older adults' needs. For walking-based locomotion, we employed the commercial device, the KAT Walk mini S \cite{katvrcom} for ensuring the olders' safety. Based on the pilot study results, we ultimately chose the SilverCycling system as the representative technique of natural motion-based technology. To assess the effectiveness of the SilverCycling system on older adults' spatial orientation and overall experience, we conducted a within-subject experiment in the open-road urban environment, employing the joystick-based controller (JC) navigation as a baseline, following Bowman's experiment framework and path-integration task \cite{bowman1998methodology, wiener2012route}. 

The user study engaged 16 participants, all aged 60 years or older. They were instructed to use both the SilverCycling system and the controller to travel through the open-road urban environment along the designated route in VR. Following this, they were asked to complete the \textit{Intersection Direction Task}, \textit{Landmark Sequence Task}, semi-structured interviews, and follow-up scales, including the Simulation Sickness Questionnaire (SSQ), the Immersive Presence Questionnaire (IPQ), and NASA Task Load Index (NASA-TLX). The results show that SilverCycling effectively improves the spatial orientation ability compared to the controller-based locomotion techniques in the open-road urban environment.
We then identified three potential factors (e.g., motion cues, familiarity) that might influence the older adult's spatial orientation in the open-road urban environment in VR. Additionally, we derived four design implications for VR locomotion techniques aimed at improving the user experience for older adults.

In summary, our work makes two contributions:
\begin{itemize}
\item We demonstrated that natural motion-based methods, such as bike-based locomotion, improve spatial orientation for older adults in open-road VR environments compared to JC in VR, as evidenced by the design and evaluation of our SilverCycling system.
\item We identified potential factors of SilverCycling that may contribute to its advantages in supporting spatial orientation; We also derived design considerations for designing locomotion techniques for older adults to sustain spatial orientation and overall user experiences.
\end{itemize}

\section{BACKGROUND AND RELATED WORK}
In this section, we introduce the importance of providing older adults with improved spatial orientation and interaction when using VR technology. We then summarize how locomotion techniques can impact users' spatial orientation and experiences, focusing on natural motion-based locomotion techniques, and explaining the rationale behind the techniques and methods.

\subsection{Older Adults and VR}
With the global population undergoing a significant demographic shift, the proportion of older adults is expected to nearly double from 12\% in 2015 to 22\% by 2050 \cite{WHOageing}. Despite their increasing numbers, older adults frequently face social and digital exclusion, exacerbated by technologies that fail to accommodate their unique needs \cite{d2020invisible, Tsetoura2022Technological}. Addressing this exclusion requires prioritizing the development of technologies like VR that are specifically tailored to their needs.

VR has emerged as a promising technology with numerous advantages for older adults. It can offer a safe and accessible platform for exploring virtual environments \cite{ortet2022cycling}, provide training and practice opportunities to maintain and improve cognitive abilities and mobility \cite{du2024lightsword, mrakic2018effects}, and encourage social interaction and entertainment \cite{xu2023designing}. Incorporating VR into older adults' lives has been proven to enhance their well-being in terms of cognitive ability, physical health, and social needs, aligning with the increasing demand for innovative solutions to support successful aging \cite{10.1145/3351235}.

However, due to differences from younger adults, older adults may face greater challenges when using VR, which could impact their ability to fully enjoy VR benefits. For instance, spatial orientation, which is crucial for navigating unfamiliar environments, often declines with age, thus impacting the ability to interact effectively with surroundings \cite{devlin2001mind, galasko1998integrated, waller2013handbook}. This decline is evident in VR spatial tasks where older adults are prone to spatial memory errors, travel longer distances to locate goals, and struggle with forming cognitive maps of the environment \cite{moffat2006age}. Additionally, Wu et al. found that compared to younger adults, older adults experience more difficulties in VR interactions such as selection and rotation \cite{Wu:2024:OlderAdults}. These findings highlight the need for VR adaptations that are further tailored to the mobility and spatial orientation capabilities of older adults.

Moreover, Kuo et al. have highlighted that reliance on turn-by-turn navigation systems can hinder users' independent wayfinding abilities \cite{10.1145/3410530.3414424}. Therefore, it is crucial to develop VR aids that are intuitive and supportive, enabling older adults to more effectively utilize their spatial orientation and mobility, rather than merely providing direct turn-by-turn directions. However, there is a significant gap in research focused on designing VR interfaces that accommodate the declining spatial and mobility capabilities of older adults, specifically aiming to enhance their spatial orientation and overall user experience.

\subsection{Spatial Orientation and VR Locomotion}
\subsubsection{VR Locomotion Techniques. }Spatial orientation refers to the ability to perceive and understand the locations, directions, and motions of objects in the surrounding environment \cite{wolbers2010determines}. Locomotion, or self-propelled movement, significantly influences users' spatial perception in VR by continually altering the visual relationship between the user and their environment \cite{plumert2007emerging, hale2014handbook, bowman1998methodology}. Disorientation and motion sickness in VR are often experienced, due to discrepancies between virtual and actual movements, complicating the maintenance of spatial orientation \cite{kim2018virtual, peruch2000transfer}. Given that older adults face greater difficulties with VR spatial tasks, it is crucial to design locomotion techniques appropriate to their ability to enhance their spatial orientation and overall experience in VR \cite{moffat2006age}.

While real walking has been recognized as the most natural method for spatial performance in VR, due to space and device limitations, it is impossible to simulate real walking experiences in VR in an open natural setting \cite{ruddle2006efficient, ruddle2009benefits, Haibach2009Egomotion}. Consequently, diverse locomotion techniques are developed.  Boletsis divided them into four types: motion-based, controller-based, teleportation-based (TP), and room scale-based \cite{boletsis2017new}. TP characterized by its non-continuous movement, allows users to move instantaneously to a pointed location, facilitating rapid movement of virtual spaces. However, the instantaneous nature of teleportation can diminish spatial awareness and navigation skills, leading to longer processing times for visual feedback and higher error time rates in tasks requiring precise spatial understanding \cite{boletsis2017new, kim2021evaluation}. The room scale-based locomotion method allows users to physically move within a designated space, but it may not be applicable to wider open space.

\textbf{Motion.} Motion-based locomotion provides continuous mobility while utilizing several forms of physical movement to facilitate interaction. It is also considered potentially beneficial for enhancing spatial orientation, as it provides body-based information that is believed to significantly improve spatial understanding \cite{bakker1999effects}. WIP is one of the prevalent motion-based locomotion methods, that allows users to simulate walking by performing a stepping motion on the spot. WIP locomotion offers cost-effectiveness and convenience, excels in simple spatial orienting tasks, and provides proprioceptive feedback akin to natural walking \cite{nilsson2016walking}. However, WIP often results in greater fatigue and effort compared to controller-based locomotion, and WIP also suffers from issues like lack of inertia, latency, and difficulties in turning, which can complicate navigation \cite{cardoso2016comparison, nabiyouni2015comparing}. ODW, which secures users with a safety harness to maintain position and allows walking or running, was selected as a potential method for our study. However, its widespread use is limited by significant installation requirements, cost, and storage space, and some believe that it still does not provide a natural movement experience \cite{bozgeyikli2016locomotion}. As mentioned earlier, RDW poses safety risks for older adults \cite{kontio2023feel}. We, therefore, include WIP and ODW in our study.

In addition to mimicking walking movements, several methods that mimic other natural movements have been developed, including bike-based, stepper-based, and rowing-based locomotion \cite{di2021locomotion, Wei2023dragon}. Bike-based locomotion in VR offers a promising approach to enhance spatial orientation and user experience, particularly for older adults. Cycling, a low-to-moderate intensity exercise, is accessible to individuals of all fitness levels, including those with little athletic experience \cite{budi2021cycling}. It provides a safe, cost-effective activity that combines the familiarity of a seated exercise with the physical benefits of moderate exertion, offering potential improvements in spatial orientation \cite{freiwald2020walking}. Studies also suggest that VR cycling improves cognitive and physical abilities \cite{anderson2012exergaming, karssemeijer2019exergaming}. As a long-established form of transportation and exercise, cycling's widespread use boosts its viability in VR, appealing to older adults who may recall cycling from earlier years \cite{oosterhuis2016cycling, winters2015grew}. Given these benefits, we incorporated bike-based locomotion into our study to further explore its impact on spatial orientation and user experience among older adults. We did not choose stepper and rowing VR locomotion methods due to their higher physical demands, safety concerns, and lower familiarity among older adults compared to bike-based locomotion. 

However, the varying levels of interaction fidelity in these systems can impose different degrees of physical and cognitive load, potentially impacting users' spatial orientation \cite{bowman1998methodology, nabiyouni2015comparing}. Consequently, there is a growing interest in body-centric locomotion that allows users to navigate virtual environments using body parts other than their feet, such as head tilt, arm swing, or torso lean, to reduce physical demand and installation cost while maintaining the benefits of body-based locomotion for obtaining information \cite{gao2021evaluation, zielasko2016evaluation, kitson2017lean}. Arm-based locomotion, such as Armswing, relies solely on the user's arm, which is analogous to a normal walking gait. This allows the user to continuously walk in the direction their body is facing, while still having the freedom to look around \cite{pai2017armswing}. However, several studies found that arm-based locomotion did not perform well in spatial awareness tasks, resulting in more turning errors \cite{wilson2016vr, pai2017armswing}. Torso-based techniques control virtual locomotion by leaning into the direction they want to travel ~\cite{10.1145/2788940.2788956, kitson2017lean}, while head-based locomotion by following the head direction ~\cite{kitson2017lean}. Although both torso-based and head-based techniques show potential for high spatial awareness and offer a seated experience for older adults and individuals with mobility disorders, body-centric locomotion is perceived as unnatural and may require higher cognitive demand \cite{ Zielasko2016, kitson2017lean, ganapathi2019elicitation, gao2021evaluation}. Additionally, four VR locomotion techniques were explored by Kontio: Supine, Rings, Dip Rack, and Chair \cite{kontio2023feel}. All required high muscle intensity and are unsuitable for older adults, although Chair was relatively easier.

\textbf{Controller.} Controller-based locomotion, particularly JC, is widely recognized as the standard method for navigating in VR. Although it may not provide the same level of spatial orientation as physical walking \cite{williams2011evaluation}, it is favored for its familiarity, ease of implementation, low cost, and safety \cite{kim2021evaluation}. Compared to alternatives like the moderate-fidelity locomotion method, controller-based locomotion as a well-designed offers better spatial awareness and requires shorter completion times while delivering comparable spatial task performance \cite{zielasko2016evaluation, kitson2017lean, Langbehn2018, bozgeyikli2019locomotion}. In our study, we employed controller-based locomotion as a baseline according to its strengths in navigation tasks.

\subsubsection{Comparative Studies of VR Locomotion Techniques on Spatial Orientation} 
To evaluate spatial orientation in VR, navigation plays a crucial role as a widely used interaction task \cite{bowman20043d}. Bowman categorized navigation tasks into two components: wayfinding and travel. Wayfinding entails the construction and upkeep of a spatio-cognitive map, which is crucial for identifying routes \cite{bowman1998methodology}. Travel, which relies on information gathered using different travel techniques, may also enhance wayfinding.

Based on the navigation tasks, Langbehn et al. evaluated three different locomotion techniques—JC, TP, and RDW—and found no significant effect on overall spatial orientation \cite{Langbehn2018}. Kim et al. compared the effects of four locomotion methods—WIP, running-in-place (RIP), JC, and TP—and found that TP and RIP were associated with higher error time ratios, indicating challenges in adapting to instant visual changes and maintaining balance, respectively. JC resulted in fewer wrong turns and collisions, suggesting it might support better continuous movement and spatial orientation during navigation \cite{kim2021evaluation}. Freiwald et al. developed an innovative VR Strider, a novel locomotion user interface (LUI) for seated VR with a rotation mechanism called anchor-turning, operated via the controller, to facilitate turning. They evaluated users’ angular and distance perception between the VR strider and JC and found that the VR strider outperformed JC \cite{freiwald2020walking}. 

In summary, previous studies have highlighted the benefits of natural motion-based VR locomotion techniques on spatial orientation, primarily focusing on younger adults. However, the specific impact of these techniques on older adults remains underexplored. To address this research gap, our study employed the aforementioned Bowman's navigation framework and path-integration tasks to assess the effects of natural motion-based locomotion methods, including WIP, ODW, and cycling, on the spatial orientation and overall user experience of older adults, with JC serving as the baseline \cite{bowman1998methodology, wiener2012route}.

\section{Pilot Study}
\begin{figure}
    \centering
    \includegraphics[width=1\textwidth]{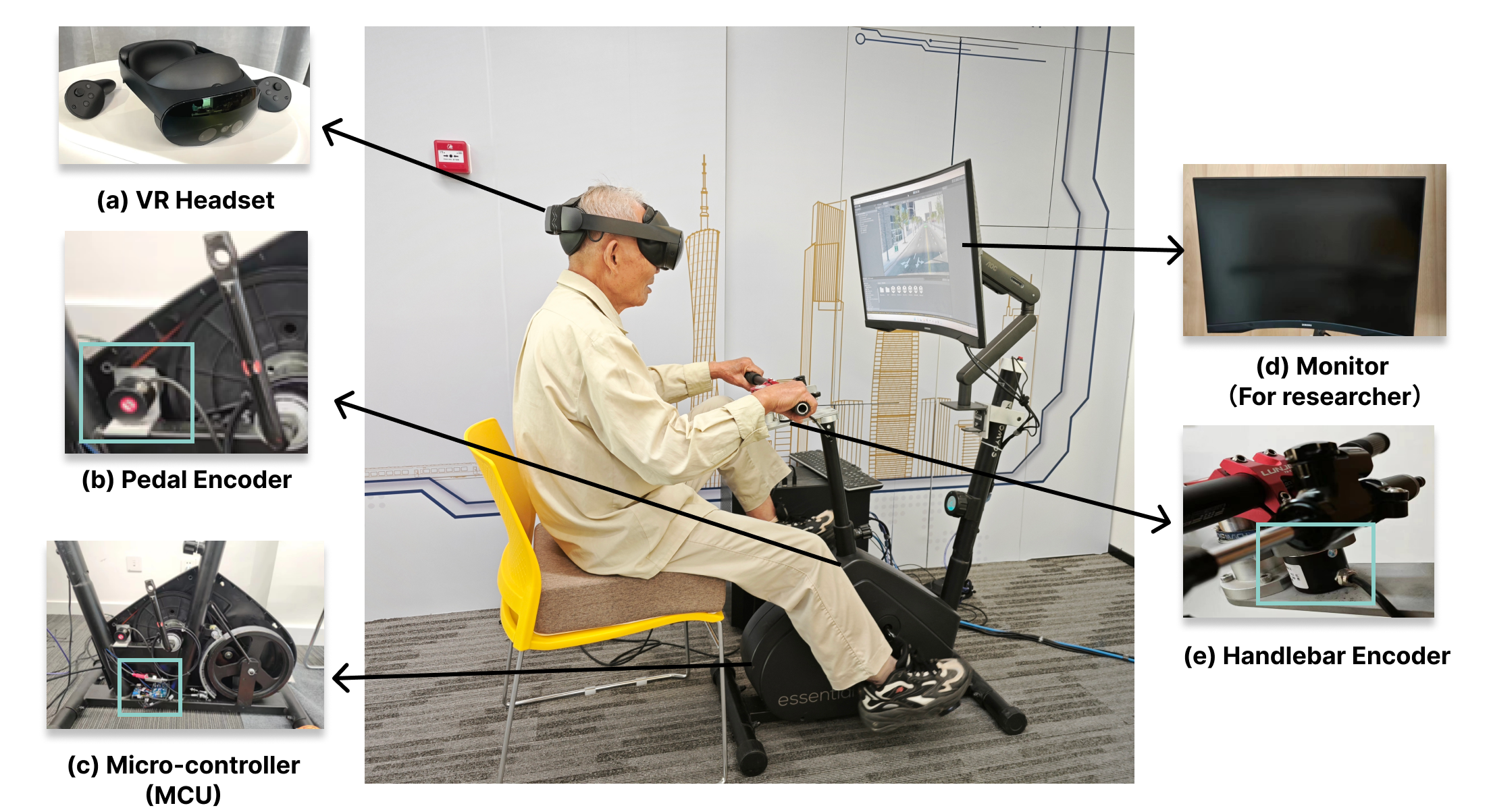}
    \Description{SilverCycling System Components and the main working process. The setup includes a Modified Indoor Cycling Machine equipped with a (e) Handlebar Encoder and a (b) Pedal Encoder to detect physical movements. (c) MCU reads input signals and subsequently transmits these signals to the PC. (a) The users wear the VR headset and movement is synchronously mapped to the camera movement within the VR environment in the headset. (d) The monitor is used for observation by researchers.}
    \caption{SilverCycling System Components and the main working process. The setup includes a Modified Indoor Cycling Machine equipped with a (e) Handlebar Encoder and a (b) Pedal Encoder to detect physical movements. (c) MCU reads input signals and subsequently transmits these signals to the PC. (a) The users wear the VR headset and movement is synchronously mapped to the camera movement within the VR environment in the headset. (d) The monitor is used for observation by researchers.}
    \label{fig:System}
\end{figure}

\begin{figure}
    \centering
    \includegraphics[width=1\textwidth]{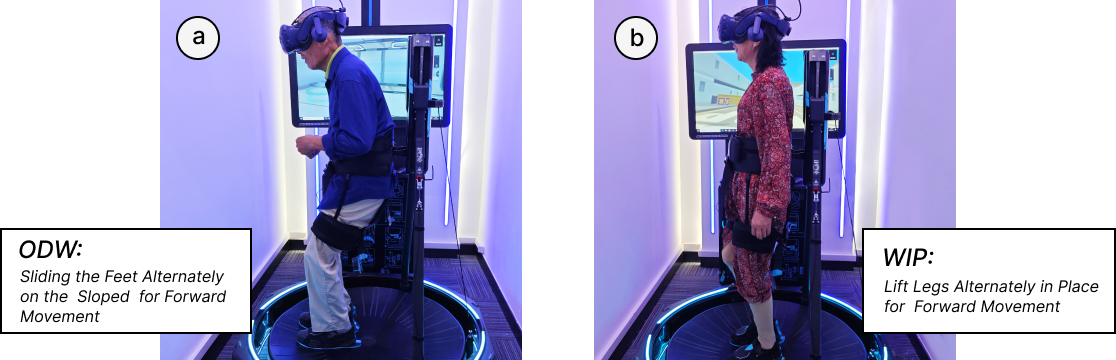}
    \caption{Illustration of two walking locomotion modes: (a) omnidirectional walking (ODW) setup and (b) walking-in-place (WIP) setup.}
    \label{fig:walking}
    \Description{Illustration of two walking locomotion modes: (a) omnidirectional walking (ODW) setup and (b) walking-in-place (WIP) setup.}
\end{figure}

Before conducting the full study, we piloted the feasibility and user acceptance of bike- and walking-based locomotion among four older adults (3 males and 1 female; mean age = 68 years, SD = 8.84). For the bike-based locomotion, as shown in Figure.\ref{fig:System}, we developed the SilverCycling System tailored to older adults’ needs (see next section for details). For the walking-based locomotion, we tested two common simulated walking modes: ODW and WIP, using a commercial device, the KAT Walk mini S equipped with a strapped safety harness, as shown in Figure.\ref{fig:walking} \cite{katvrcom}. Each participant tested the SilverCycling System, ODW, and WIP in separate sessions lasting between 5 to 10 minutes, focusing on forward movement and turning.

The results revealed several challenges with ODW: three participants required at least five minutes to become accustomed, while one struggled to understand and effectively use the system, failing to achieve forward movement. Additionally, participants noted that ODW required substantial effort and led to fatigue, primarily due to the unnatural sliding motion, and effective movement necessitated leaning to counteract the strap and the inertia of the rotating axis. Some of these problems were also mentioned in prior work \cite{bozgeyikli2016locomotion}. While WIP was easier to learn and use than ODW, participants still found it tiring and monotonous, lacking resistance (haptic feedback) and presenting challenges in controlling body rotation due to the inertia of the rotating axis. The feedback provided was consistent with findings from earlier research’s findings \cite{cardoso2016comparison, nabiyouni2015comparing}. In contrast, all participants found the SilverCycling System was the easiest to learn and use, reporting that it felt natural and less exhausting. However, the only drawback noted was that turning on the bike induced more dizziness compared to walking, attributed to larger rotational movements.

Based on the results, we chose the SilverCycling System as the representative natural motion-based locomotion method for our study, aiming to further explore its impact on enhancing spatial orientation and overall user experiences for older adults without diminishing their willingness to participate.

\section{SilverCycling SYSTEM DESIGN}
In this section, we first detail the design considerations inspired by prior work. Subsequently, we introduce the SilverCycling System, designed to provide older adults with a secure and user-friendly locomotion technique that seamlessly integrates with their movements in the VR environment, with the overarching goal of improving spatial orientation.

\subsection{Design Considerations}
\textbf{DC1: Provide Safe and Comfortable VR Experience for Older Adults.} In designing cycling locomotion systems for older adults, our primary goal was to balance safety with realism and provide a comfortable experience. Inspired by Matviienko et al.'s research, which suggests a 'bikeless' setup increases perceived safety, we adopted a customized indoor cycling machine instead of a real bike \cite{matviienko2023does}. This design choice is in line with McGlynn and Rogers' findings that older adults prefer seated virtual experiences \cite{mcglynn2017design}. We carefully adjusted the seat, handlebars, and pedals, focusing on accessibility and comfort, to enhance the safety and enjoyment of our participants. These modifications ensure the VR experience is both secure and immersive for older adults.

\textbf{DC2: Provide Natural and Intuitive Control Mechanism.} Effective bike simulators hinge on precise control mechanisms, primarily steering and pedaling, and the control over steering and pedaling leads to the highest level of cycling control \cite{matviienko2023does}. Consequently, our design incorporates both steering and pedaling functionalities. 

In designing steering mechanisms for older adults, existing control methods utilizing buttons and platform tilting are not ideal. For instance, the steering method proposed by Katsigiannis et al.involving buttons on either side of the handlebar, diverges significantly from the real biking experience \cite{katsigiannis2018qoe}. This method may not be intuitive or natural for older adults accustomed to traditional cycling. Additionally, while the tilting of simulating platforms mimics the upper body movements of cyclists, providing some degree of realism for users, the complexity and cost of these setups are high \cite{shoman2021bicycle}. For older adults, such complex equipment might be challenging to operate and maintain. Handlebar rotation is achieved through either the free movement of the handlebar with the front wheel on a floor or a turntable, or by using a front wheel with a front-mounted fork and a movable handlebar \cite{matviienko2018augmenting, matviienko2022baby, lieze2020using, ullmann2020bikevr}. In most cases, a 10° leftward handlebar rotation corresponds to a 10° shift in the simulation view, mirroring real-world cycling dynamics. Informed by existing research, our steering design for older adults employed handlebar rotation, offering an intuitive approach that minimizes motion sickness while preserving the natural steering interaction and facilitating easier adaptation for older users.

\textbf{DC3: Provide Immersive and Visual Experience.} Research supports the use of HMDs for activities requiring frequent turns and spatial orientation. Bowman et al. highlight HMDs as a suitable option for such tasks \cite{bowman2002empirical}. Further, Lieze et al. indicate that older users experience a more heightened sense of reality using VR headsets compared to traditional 3D-CAVE setups \cite{lieze2020using}. Informed by these findings, we selected HMDs as the primary display for our SilverCycling system, aiming to optimize the visual experience for older adults. 

\subsection{Overview of System}
Based on the aforementioned design considerations, we developed the SilverCycling system. Figure.\ref{fig:System} shows the main system components and their functions. The system consists of three main components:

1. A modified indoor cycling machine, including handlebars, pedals, and a seat for easy access by older adults. 

2. An integrated input capture and transmission system, consists of an absolute handlebar encoder and an incremental pedal encoder, an Arduino-compatible microcontroller Unit (MCU), and a WIFI router. 

3. A display system, including a PC, a monitor, and a set of VR headsets. 

\textbf{Forward Mechanisms.} We utilized a rotary encoder located near the pedals to capture the user's inputs related to forward and backward movements. This input, represented in rounds per second, is then subject to a clamped linear mapping, effectively constraining it within the interval of [-1, 1]. This mapping facilitates subsequent customization and further system functionality. After processing, we employed the processed data to control the camera's forward velocity within VR environments.

\textbf{Turning Mechanisms.} We employed an absolute rotary encoder near the handlebar of the SilverCycling system to capture participants' turning inputs, providing rotation degree information. Positive values represent right turns, while negative values correspond to left turns. Following this, we applied a clamped linear mapping to the input, restricting it to the range of [-1, 1], allowing for further customization. We utilized the processed data to calculate and control the camera's rotation within the VR environment.

\subsection{Speed of SilverCycling System}
After signing the consent form, we invited five older adults (4 females and 1 male; mean age = 61 years, SD = 1.50) to engage with the speed chosen. Participants were required to test various forward and turning speeds within the VR environment. In the study, participants were exposed to varying forward speeds, increasing in 5 km/h increments from a starting point of 5 km/h, up to a maximum of 20 km/h. Additionally, they experienced turning speeds ranging from 5 to 20 degrees, divided into distinct 5-degree intervals. After each trial, participants were asked to classify the experienced speed as either too slow, too fast, or moderate. Each trial lasted approximately one minute and was conducted three times for consistency. 

We found that a forward speed of 10 km/h and a turning speed of 10 degrees per second were optimal for comfort. Additionally, we decided to remove the backward movement function from our SilverCycling system. This decision was based on observations that backward motion can easily induce motion sickness in participants, which also aligns with Flynn's suggestion \cite{flynn2003developing}. 

\section{USER STUDY}
This study aims to investigate and evaluate the effectiveness of the SilverCycling system in enhancing older adults' \textit{Spatial Orientation} and other user experience, including \textit{Motion Sickness}, \textit{Presence}, \textit{Perceived Workload}, \textit{Subjective Level of Enjoyment, Safety, Comfort},  while traveling in the open-road urban city. 

\subsection{Study Design}
The study was designed to be within-subject with two conditions: SilverCycling and JC. We selected JC as the baseline technique, which is a common locomotion technique known for its ease of control and improved spatial orientation \cite{kitson2017lean}. Then, we designed a user study based on a path-integration task, specifically the \textit{Intersection Direction Task} and \textit{Landmark Sequence Task}, as utilized in prior research for assessing the SilverCycling system performance \cite{bowman1998methodology, wiener2012route}. Additionally, we employed metrics similar to those used in previous studies, including \textit{Spatial Orientation}, \textit{Motion Sickness}, \textit{Presence}, \textit{Perceived Workload}, \textit{Subjective Level of Enjoyment, Safety, Comfort}, to evaluate the effectiveness of these two locomotion techniques \cite{bowman1998methodology, matviienko2023does}. To enhance the internal validity of the study by minimizing order-related biases and potential learning effects, we first created two distinct routes with the same difficulty level, and then we used a balanced Latin square to counterbalance the sequences of routes and locomotion modes.

Both SilverCycling and JC conditions were explored within a controlled laboratory environment. Participants were assigned the tasks of both SilverCycling and JC along a task route in VR under all controlled experimental conditions. Each task took around 5 minutes.

\subsection{Participants}
We recruited 16 participants, all of whom met the following criteria: 1) aged 60 or older, 2) having fully functional limbs, 3) with normal or corrected-to-normal vision, and 4) without cardiovascular or other serious diseases. Table \ref{table:demographic} shows the detailed demographic information of the participants. Among the 16 participants, 11 of them are female, and 5 are male, with an average age of 64 years old (SD=6.27). Participants reported varying proficiency levels in cycling (M=2.06, SD=0.83) and JC usage (M=1.38, SD=0.70). They also reported diverse frequencies of using cycling (ranging from never used to once a week) and JC in the past two years (ranging from never used to less than once a month).
\begin{table}[]
\begin{tabular}{|c|c|c|c|c|c|c|c|}
\hline
ID &
  Age &
  Sex &
  Education &
  \multicolumn{1}{c|}{\begin{tabular}[c]{@{}c@{}}Cycling \\ Proficiency\end{tabular}} &
  \multicolumn{1}{c|}{\begin{tabular}[c]{@{}c@{}}JC \\ Proficiency\end{tabular}} &
  \multicolumn{1}{c|}{\begin{tabular}[c]{@{}c@{}}Cycling \\ Frequency\end{tabular}} &
  \multicolumn{1}{c|}{\begin{tabular}[c]{@{}c@{}}JC \\  Usage Frequency\end{tabular}} \\ \hline
1  & 60 & Female & College     & 3 & 3 & 5 & 2 \\ \hline
2  & 64 & Female & College     & 3 & 1 & 2 & 2 \\ \hline
3  & 63 & Female & College     & 3 & 1 & 3 & 2 \\ \hline
4  & 84 & Male   & College     & 3 & 1 & 3 & 2 \\ \hline
5  & 60 & Female & College     & 2 & 2 & 1 & 1 \\ \hline
6  & 61 & Female & College     & 1 & 1 & 2 & 2 \\ \hline
7  & 62 & Male   & College     & 2 & 1 & 1 & 1 \\ \hline
8  & 61 & Female & High School & 3 & 3 & 2 & 2 \\ \hline
9  & 73 & Male   & College     & 2 & 1 & 1 & 1 \\ \hline
10 & 67 & Male   & High School & 1 & 1 & 1 & 1 \\ \hline
11 & 60 & Female & High School & 1 & 1 & 1 & 1 \\ \hline
12 & 60 & Female & College     & 1 & 1 & 1 & 1 \\ \hline
13 & 60 & Female & College     & 2 & 2 & 4 & 1 \\ \hline
14 & 61 & Female & Master      & 2 & 1 & 2 & 2 \\ \hline
15 & 60 & Female & College     & 3 & 1 & 4 & 2 \\ \hline
16 & 60 & Male   & College     & 1 & 1 & 2 & 2 \\ \hline
\end{tabular}
\vspace{10 pt}
\caption{Participants' Demographic Information. The table provides demographic details of the study participants, including their age, gender, education level, and proficiency with cycling (rated on a scale of 1 to 3, from novice to proficient), proficiency with joystick-based controller (JC) (also rated on a scale of 1 to 3, from novice to proficient), frequency of cycling (rated on a scale of 1 to 5, from never to daily), and frequency of joystick-based controller usage (rated on a scale of 1 to 5, from never to daily) over the past two years}. 
\label{table:demographic} 
\end{table}

\subsection{Apparatus}
\subsubsection{SilverCycling system. }The setup, detailed previously and depicted in Figure \ref{fig:System}, includes a modified indoor cycling machine designed for older adults. It contains a modified indoor cycling machine for older adults, enhancing the seat, handlebars, and pedals for accessibility. The system utilizes handlebar and pedal encoders, processed by an MCU and relayed as TCP/IP packets to a PC via a WiFi router for controlling the movement in 3D environments. For an immersive user experience, a VR headset connected to the PC and an auxiliary screen provides real-time visualization for researchers. Users navigate the environment using the handlebar to adjust turning speed and angle while pedaling influences forward movement, as shown in Figure.\ref{fig:setup}a. The system limits the maximum forward speed to 10 km/h and turning speed to 10 degrees, optimizing the user's experience.

\subsubsection{Joystick-based Controller (JC)} 
\begin{figure}
    \centering
    \includegraphics[width=1\textwidth]{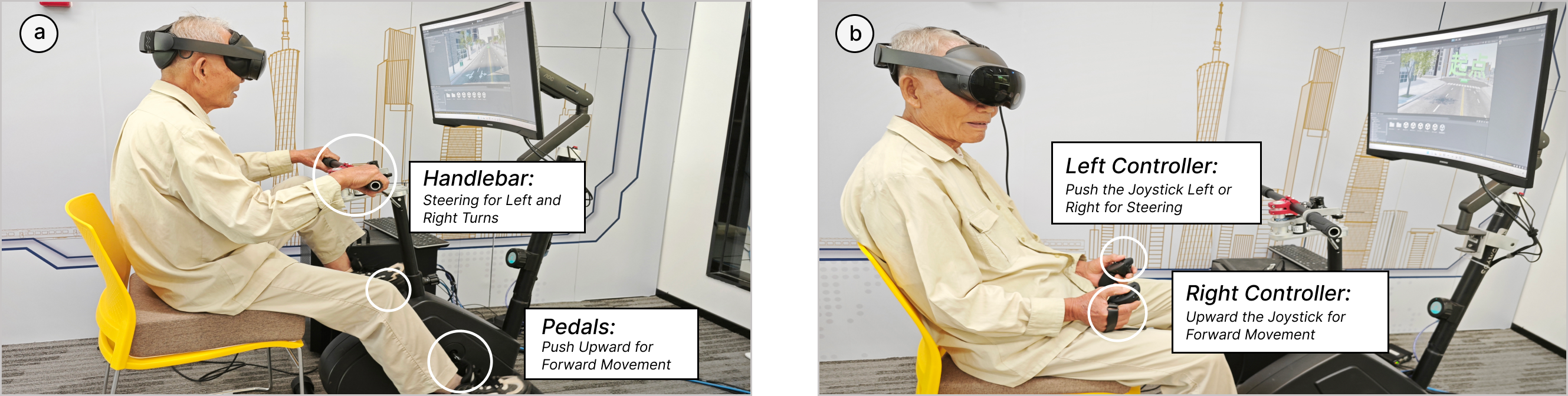}
    \caption{The SilverCycling system and JC. (a) Depicts the interaction method of the SilverCycling system. (b) Shows the interaction method of the.}
    \label{fig:setup}
\end{figure}

The setup, utilizes the same PC, VR headset, and cable in the SilverCycling system, This setup uses JC of Meta Quest Pro to navigate within 3D environments \cite{2023Meta}. Drawing insights from our pilot study, we refined the control scheme to mitigate motion sickness. A single joystick manages both forward movement and turning, which can be disorienting. To enhance user comfort, the setup assigns the left controller's joystick for steering (left or right) and the right controller's joystick exclusively for forward movement, as shown in Figure.\ref{fig:setup}b. Additionally, to minimize differences from the SilverCycling system while simultaneously enhancing the user experience, the setup also caps the maximum forward speed at 10 km/h and the turning speed at 10 degrees. 

\subsubsection{VR software and hardware}
For all the setups, we used the same computer with 48 GB of RAM, a Nvidia GeForce GTX 4090 graphics card and an Intel Core i7-7700HQ7 processor with a base frequency of 2.80 GHz. To present the VR environment to the participants, we employed a Meta Quest Pro as the Head-mounted display (HMD). The glasses have 10 advanced sensors to support 6 degrees of freedom inside-out simultaneous localization and mapping tracking. To implement the virtual environment, we used the Unity engine version 2021.3.27f1. For rendering the virtual scene into the HMD, we used a streaming cable. To document the study process, we utilized both the phone camera and ShareX, a screen recording software \cite{2023ShareX}.

\begin{figure}
    \centering
    \includegraphics[width=1\textwidth]{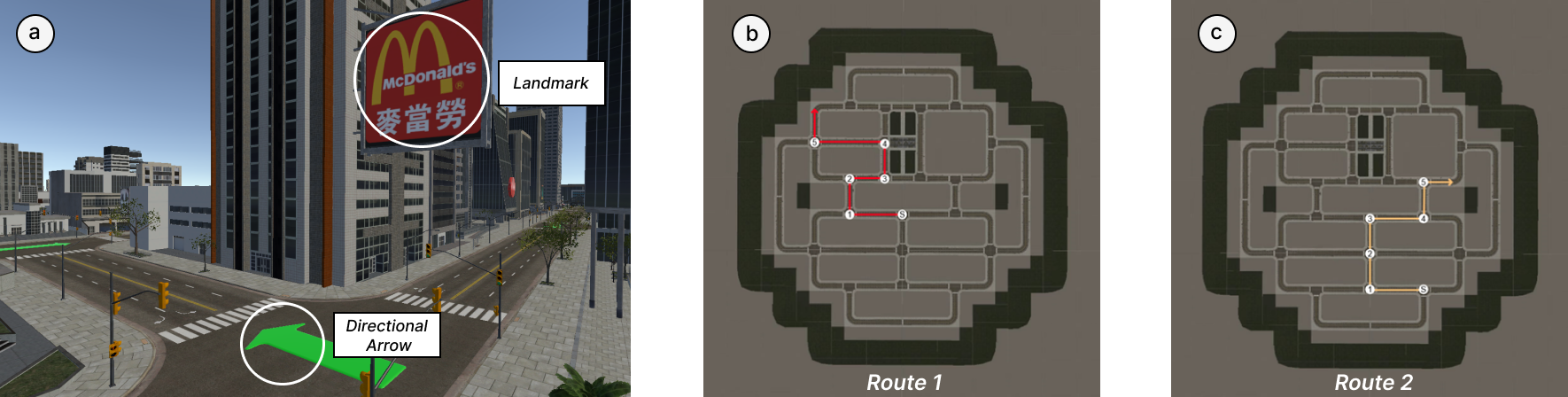}
    \caption{Task Scene and Routes in a Medium-Scale Urban Environment. (a) Depicts the task scene created using the Fantastic City Generator \cite{2023City}, showcasing a medium-scale urban setting with a unique single landmark on a building near selected intersections. (b) Represents Route 1, and (c) illustrates Route 2, both designed to avoid potential learning effects. Each route comprises five distinct intersections, each marked by a unique landmark, and directional guidance is provided by green arrows on the ground.}
    \label{fig:scene}
\end{figure}

Figure \ref{fig:scene} shows the route design of the tasks. We employed Fantastic City Generator to construct a medium-scale urban environment, incorporating a randomized placement of a unique single landmark on a building near the selected intersections \cite{2023City}. We established two different routes to mitigate the potential learning effect. Taking into account considerations of working memory capacity and experimental complexity, our study eventually set up five distinct intersections, each marked by a unique landmark related to older adults' daily life, for each route \cite{unsworth2005automated, cowan2001magical}. Meanwhile, each interaction was indicated by a green arrow on the ground to indicate the direction.

\subsection{Measurements}
To investigate and evaluate the effectiveness of SilverCycling in spatial orientation and user experiences, we measured the following dependent variables: 
\indent
\begin{itemize}
\item \textit{Spatial Orientation}: To provide a comprehensive measure of spatial orientation, we employed two tasks, \textit{Intersection Direction Task} and \textit{Landmark Sequence Task}. The measure of The \textit{Intersection Direction Task} assesses the ability to maintain and update directional information during navigation. Participants need to draw the entire experiment route consisting of five intersections on the paper with blank cells marked with a starting direction. With three possible movement directions and an equal number of trials requiring left, right, and straight responses. Meanwhile, \textit{Landmark Sequence Task} evaluates the recognition and sequential recall of landmarks. It consists of a set of five landmark-based single-choice questions. Each question offered four options, among which only one was correct. Participants are required to sequentially select the landmarks that correspond to the route they have navigated. 
\item \textit{Motion Sickness}: After each setup, participants completed the SSQ to evaluate their overall level of motion sickness \cite{kennedy1993simulator}. The SSQ score was computed using the formula \cite{kennedy1993simulator}. Total SSQ scores ranging from 20 to 30 indicate minimal to moderate motion sickness, while scores exceeding 40 suggest an unfavorable simulator experience \cite{caserman2021cybersickness}. 
\item \textit{Presence}: for each set-up, every participant evaluated their sense of presence within the virtual environment by utilizing IPQ \cite{schubert2003sense}. The IPQ encompasses various subscales, each rated on a scale from 1 (lowest) to 7 (highest). These subscales include spatial presence, involvement, and the degree of experienced realism.
\item \textit{Perceived Workload}: After each condition, participants accessed the workload by NASA-TLX \cite{hart2006nasa}. The NASA-TLX comprises six subscales, each rated on a scale ranging from 1 (lowest) to 7 (highest), except the performance, where a score of 1 indicates optimal performance.
\item \textit{Subjective Level of Enjoyment, Safety, Comfort}: for each locomotion, we asked participants to report their subjective level of enjoyment, safety and comfort using a 7-point scale (1 – strongly disagree or low sense of enjoyment/safety/comfort, 7 – strongly agree or high sense of enjoyment/safety/comfort) for the following statements: 
(1) ``I found this cycling/JC experience interesting'', (2)``I found this cycling/JC experience safe'', and (3) ``I found this cycling/JC experience comfortable''.

\end{itemize}
\subsection{Procedure}
\begin{figure}[h]
    \centering
    \includegraphics[width=1\textwidth]{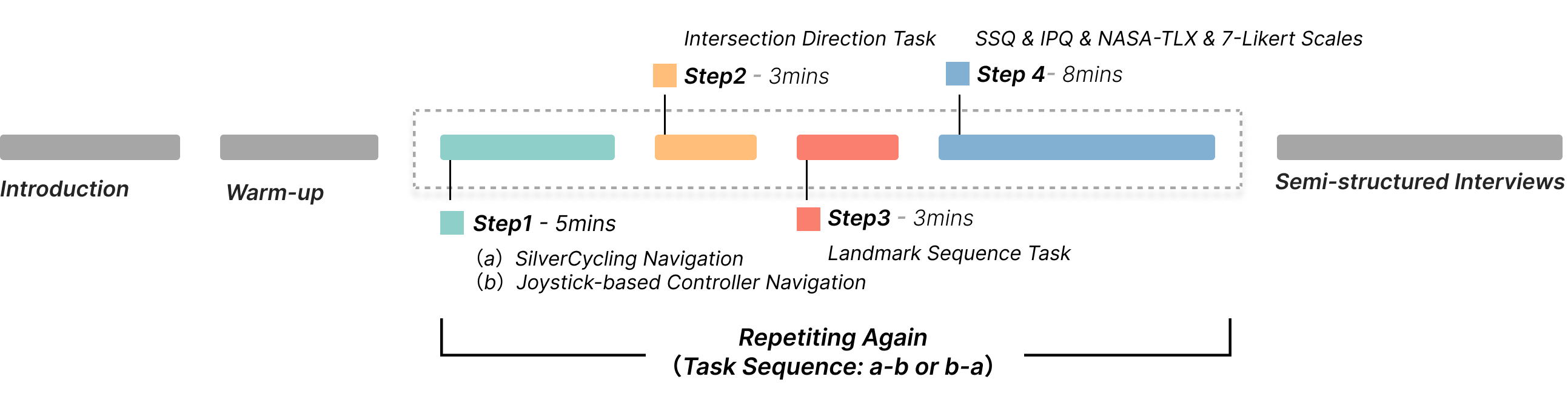}
    \caption{Overview of the experimental flow, from introduction and warm-up to semi-structured interviews, detailing a sequence of tasks with specified durations. Initial tasks include SilverCycling and JC locomotion (Step 1), followed by the Intersection Direction Task (Step 2), the Landmark Sequence Task (Step 3), and evaluations using SSQ, IPQ, NASA-TLX, and 7-Likert Scales (Step 4). The sequence concludes with a counterbalanced repetition (a-b or b-a).}
    \label{fig:procedure}
\end{figure}

After obtaining informed consent, we collected participants’ demographic data. Subsequently, we provided an introductory video that provided a concise overview of the study procedures, detailed explanations of the setup types, and the tasks that needed to be completed. Participants then had the opportunity to become acquainted with both the SilverCycling and JC during a preliminary warm-up session.
The warm-up sessions included scenes, routes, and landmarks similar to those in the formal study, but with some differences. Participants were instructed to navigate along the directional arrows using the provided setups. Each warm-up session lasted around 5 minutes, concluding with participants being able to move forward and turn on independently. Once they felt comfortable with the equipment, the experiment commenced. Participants were tasked with navigating a designated route, guided by directional arrows, using either the SilverCycling or JC. Their objective was to memorize the orientation and corresponding landmark of each intersection in sequential order. Following each setup's locomotion, participants completed the \textit{Intersection Direction Task}, \textit{Landmark Sequence Task}, SSQ, IPQ, and NASA-TLX. Additionally, they reported their subjective level of enjoyment, safety, and comfort. At the end of the study, we conducted semi-structured interviews to gather the participants' feedback on spatial orientation, preferences, challenges, and suggestions regarding each setup. Each route took approximately 5 minutes to complete, and the entire study had a duration of approximately one and a half hours. Figure \ref{fig:procedure} shows the entire experiment procedure. The study was conducted with approval from the ethical review board at our university. Each participant had the right to stop and quit the experiment whenever they wanted. After completing the study, each participant received compensation according to the local standard. 

\subsection{Data Analysis}
For the quantitative data, one researcher first collected and organized all the data into Excel and another researcher examined the data. The results from the \textit{Intersection Direction Task} and the \textit{Landmark Sequence Task} were converted into binary (0/1) data. Then, we employed IBM SPSS Statistics 24.0 to conduct statistical analysis \cite{2023IBM}. Given the non-normal distribution of the data from the tasks, including the SSQ data, the NASA-TLX scores, and responses from the 7-point Likert scale, we conducted the Wilcoxon Signed-Rank test for the analysis. For the IPQ data which conformed to a normal distribution, we used the paired sample T-test to analyze. Additionally, we conducted Spearman rank correlation analysis to explore the relationship between SSQ questionnaire scores and the results of the \textit{Intersection Direction Task} and the \textit{Landmark Sequence Task}. 

For the observational and interview data, researchers initially transcribed and translated all recordings using the Feishu Platform, and subsequently conducted a thematic analysis approach of the data \cite{vaismoradi2013content, 2023Feishu}. Then, two researchers independently reviewed the videos and transcriptions to become familiar with the content, which was followed by an open coding process to identify patterns and themes within the data. Any discrepancies in coding were resolved through discussion and a re-examination of the observation and interview transcripts, ensuring mutual understanding. The themes were conclusively finalized after a second iterative review.  

\section{RESULTS}
According to the data analysis, participants using SilverCycling locomotion exhibited significantly greater accuracy in \textit{Intersection Direction Task}, along with improved performance in the \textit{Landmark Sequence Task}. We also noted variances in motion when participants utilized the two locomotion techniques. Although no significant differences were observed in motion sickness and presence levels, there was a trend indicating reduced motion sickness and elevated presence levels among these participants. Additionally, we also uncovered the relationship between motion sickness and task performance. Furthermore, a majority of participants reported heightened enjoyment, safety, and comfort in comparison to employing the joystick-based controller.

\subsection{Spatial Orientation}
\begin{figure}[h]
    \centering
    \includegraphics[width=1\textwidth]{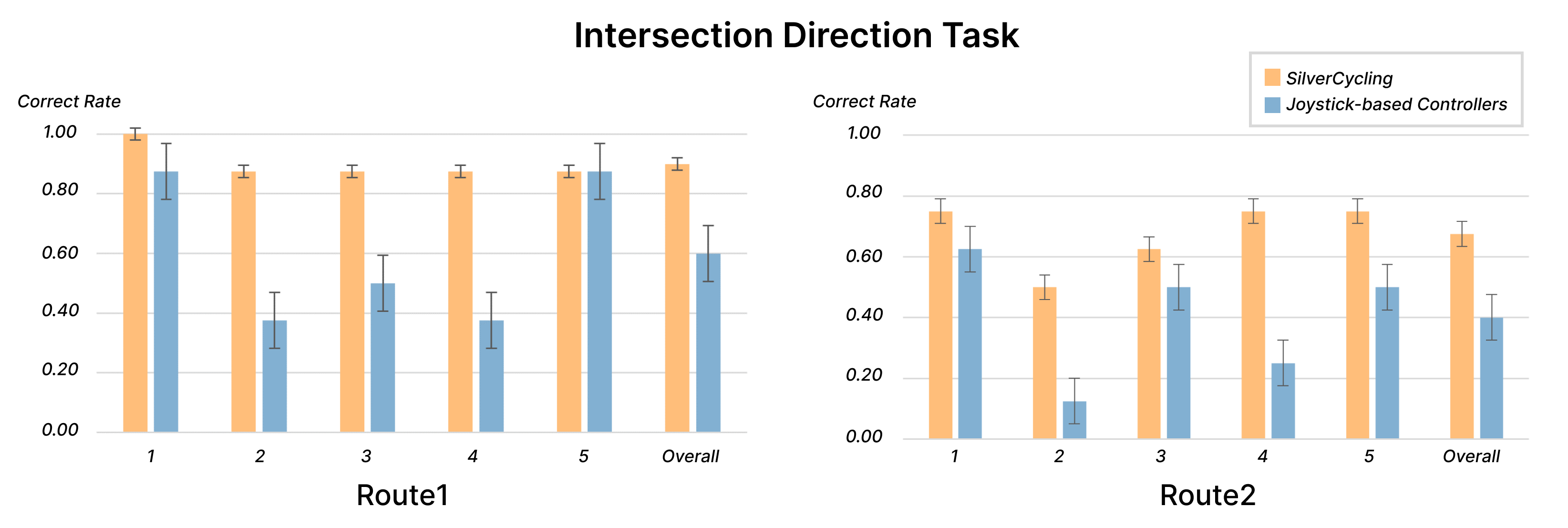}
    \caption{Correct Rates in \textit{Intersection Direction Task}. The figure (left) shows the correct rates for Route 1 in the \textit{Intersection Direction Task}. The figure (right) depicts the correct Rates for Route 2 in the \textit{Intersection Direction Task}.}
    \label{fig:inter}
\end{figure}

\textbf{Intersection Direction Task Performance.} Statistical analysis revealed a significant difference in performance on the \textit{Intersection Direction Task} between using the SilverCycling and the controller ($Z=-2.28 $, $p<0.05$). Specifically, participants using SilverCycling demonstrated a markedly higher average accuracy rate of 79\% ($SD=0.8$), compared to 50\% ($SD=2.0$) for the controller. This was further substantiated by our calculations of the average accuracy at each intersection for the routes taken by participants, which indicated that SilverCycling matched or surpassed joystick-based controller in all cases, as detailed in Figure \ref{fig:inter}.  

\begin{figure}[h]
    \centering
    \includegraphics[width=1\textwidth]{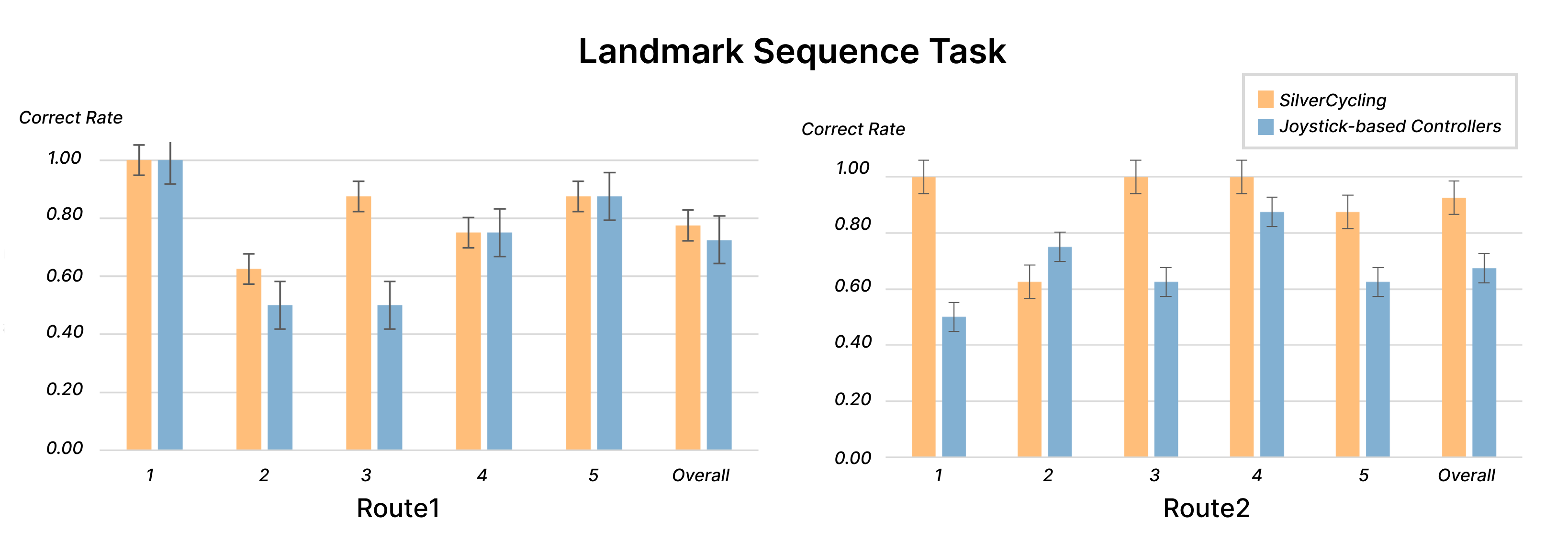}
    \caption{Correct Rates in the \textit{Landmark Sequence Task}. The figure (left) shows the correct rates for Route 1 in the \textit{Landmark Sequence Task}. The figure (right) depicts the correct Rates for Route 2 in the \textit{Landmark Sequence Task}.}
    \label{fig:land}
\end{figure}

\textbf{Landmark Sequence Task Performance.} Although the overall performance between SilverCycling and joystick-based controller showed no statistically significant differences ($Z=-1.43$, $p>0.05$) in \textit{Landmark Sequence Task}, a notable trend was observed, where the average accuracy rate for participants using SilverCycling was higher at 85\% ($SD=1.0$), compared to 70\% ($SD=2.5$) when using the controller. This trend aligns with the observations from the \textit{Intersection Direction Task}. Detailed analysis of the average accuracy at each intersection for the routes taken by participants revealed that SilverCycling either matched or exceeded joystick-based controller in almost all instances, as shown in Figure \ref{fig:land}. Additionally, it was observed that the average accuracy in the straightforward intersection was lower than the overall average for both SilverCycling and the joystick-based controller. 

\textbf{Subjective for Perceived Spatial Orientation.} We found that participants reported higher spatial orientation and perception when using SilverCycling locomotion compared to the joystick-based controller based on the statistical results ($Z=-2.88$, $p<0.05$). The interview results revealed diverse attitudes towards spatial perception with VR locomotion techniques. A majority of participants ($N=8$) reported that the SilverCycling system enhanced spatial perception, attributing this to factors like physical engagement and familiarity. For instance, P4 stated, \textit{``I believe it depends on familiarity; the more familiar, the better to remember the orientation and information. I've been riding for decades, so I can notice and remember the surroundings information.''} In terms of the physical movement, P10 expressed, \textit{``When riding, a larger body rotation should be more helpful for perceiving spatial orientation.''} Several participants ($N=4$) believed the locomotion techniques would not affect spatial perception, suggesting a neutral stance. Although one participant believed that cycling would demand more attention than using the controller, another participant felt more familiar with cycling. Some participants ($N=2$) held mixed opinions, recognizing no effect from the SilverCycling system but a negative impact on spatial perception from controller use, primarily due to the increased concentration required for directional control. As P14 expressed, \textit{``I don't think cycling contributes much to enhancing spatial memory. But the controller might even diminish spatial perception because I need to pay more attention to operating the controller.''} A minority ($N=2$) suggested joystick-based controller as potentially more effective for spatial perception in VR, emphasizing its precision and reduced physical demand. \textit{``Personally, I think it's more effective to use the controller when retracing a route without arrow indicators and relying on memory. The controller doesn't involve foot movement, which helps avoid distractions. Its clear purpose makes it the preferred choice for this task, while cycling is seen as more casual.''}, explained by P7.

\subsection {Turning Movements.} 
When observing participants' turning performance, we focused on two aspects: turning speed and turning continuity. We categorized turning speed based on the time required for a turn (t), distinguishing between 1) Fast turns (t < 10s); 2) Medium-speed turns (10s < t < 15s); 3) Slow turns (t > 15s). In addition, we categorized turning continuity into continuous turning and non-continuous turning based on whether participants had paused during their turning process.
We found that when using the SilverCycling system, the majority of participants preferred turning at medium speeds or above (N=12). Additionally, participants tended to favor continuous turning (N=14). In the case of using the controller, most participants also showed a preference for turning at medium speeds or above (N=12). In comparison to the usage of the SilverCycling system, the number of participants opting for continuous turning decreased (N=5), while those choosing non-continuous turning increased (N=11).
. 
\subsection{Head Movements. }
When observing participants' head movements, we centered on two aspects: the magnitude of head oscillations and the methods employed during landmark exploration, including lateral gaze and unidirectional gaze \cite{dunne1987cerebellar}. The magnitude of head oscillations was categorized based on the average deviation (y) from the screen's initial center along the Y-axis (perpendicular to the ground), distinguishing between 1) Large oscillations (y > 4cm); 2) Medium oscillations (2cm < y < 4cm); 3) Small oscillations (y < 2cm). We found that when using the SilverCycling system, the majority of participants exhibited either medium or large head oscillations (N=11). Participants using unidirectional gaze (N=9) and lateral gaze (N=7) were approximately equal. Notably, participants employing lateral gaze showed a preference for small-magnitude scanning (N=5).In the case of using the controller, the majority of participants exhibited either medium or small head oscillations (N=14). Additionally, participants showed a strong preference for using unidirectional gaze to search landmarks (N=14).

\subsection{Motion Sickness and Presence. } 
\textbf{Motion Sickness. }We found that both SilverCycling and joystick-based controller induce a comparably low level of motion sickness based on the SSQ scores. This was shown by non-significant effects on the overall SSQ score ($Z=-0.90$, $p>0.05$) and the sub-score of nausea ($Z=-1.10$, $p>0.05$), disorientation ($Z=-0.43$, $p>0.05$), and oculomotor ($Z=-1.4$, $p>0.05$). However, we observed a tendency for lower SSQ scores for the SilverCycling locomotion. Particularly, SilverCycling locomotion's total SSQ score ($M=7.5, SD=9.0$) was lower than JC ($M=9.8, SD=9.4$). Similar tendencies were also observed for the sub-scores of SSQ. SilverCycling locomotion showed lower scores in all three sub-scores of SSQ, nausea, oculomotor, and disorientation, as shown in Table \ref{tab:SSQ}. 

Participants reported experience of motion sickness induced by the two locomotion methods. Concerning JC, participants (N=8) reported experiencing dizziness. This was particularly noted during abrupt stops (N=2) and while adjusting the controller's turning (N=6). Several participants (N=4) described dizziness when engaging in continuous steering combined with forward motion, expressing a preference for turning without moving forward. In contrast, some participants found that rotation alone was dizzying. Some of them (N=3) reported motion sickness due to the incorrect use of the left and right functions on the controller. Additionally, some (N=2) noted that high turning speeds could induce motion sickness. Regarding the SilverCycling system, a smaller group (N=2) reported dizziness specifically during turning maneuvers. They mentioned that the turning feeling did not match reality.

\textbf{Correlation Between Motion Sickness and Task Performance. }In addition, we explored the relationship between motion sickness, visual perception, and spatial orientation. We first conducted Spearman rank correlation analysis to examine the relationship between participants' SSQ scores (Nausea/Oculomotor/Disorientation) during the use of the SilverCycling system and the controller, and their accuracy in \textit{Landmark Sequence Task} and \textit{Intersection Direction Task}. The results showed that, when using the SilverCycling system, the accuracy of the \textit{Landmark Sequence Task} had no significant correlation with SSQ, while the accuracy of the Intersection Direction Task was significantly positively correlated with all three dimensions of SSQ (Nausea: $R=0.88$, $p<0.05$; Oculomotor: $R=0.75$, $p<0.05$; Disorientation: $R=0.91$, $p<0.05$). On the other hand, when using the controller, the accuracy of the \textit{Landmark Sequence Task} was significantly positively correlated with the Oculomotor and Disorientation dimensions of SSQ (Oculomotor: $R=0.55$, $p<0.05$; Disorientation: $R=0.64$, $p<0.05$), while the accuracy of the Intersection Direction Task showed no significant relationship with any of the SSQ dimensions.

\textbf{Presence.}We calculated the IPQ scores for the general presence and its subscales, which are shown in Table \ref{tab:SSQ}. Our analysis has shown that the SilverCycling setup of presence in the virtual environments, given no statistically significant differences ($t=1.00$, $df=15$, $p>0.05$) for the general presence and all subscales, including spatial presence ($t=2.16$, $df=15$, $p>0.05$), involvement ($t=1.98$, $df=15$, $p>0.05$) and experienced realism ($t=2.11$, $df=15$, $p>0.05$). However, similar to the SSQ scores, we observed a tendency for a minor higher level of presence with JC. 

\begin{table}[h]
\centering
\sisetup{
  table-column-width=12mm, 
  table-align-text-post=false 
}
\begin{tabular}{
  l
  |S[table-format=2.1, table-column-width=3.8mm]
  S[table-format=2.1, table-column-width=3.8mm]
  S[table-format=2.1, table-column-width=3.8mm]
  S[table-format=2.1, table-column-width=3.8mm]
  S[table-format=2.1, table-column-width=3.8mm]
  S[table-format=2.1, table-column-width=3.8mm]
  S[table-format=2.1, table-column-width=3.8mm]
  S[table-format=2.1, table-column-width=3.8mm]
  |S[table-format=2.1, table-column-width=3.8mm]
  S[table-format=2.1, table-column-width=3.8mm]
  S[table-format=2.1, table-column-width=3.8mm]
  S[table-format=2.1, table-column-width=3.8mm]
  S[table-format=2.1, table-column-width=3.8mm]
  S[table-format=2.1, table-column-width=3.8mm]
  S[table-format=2.1, table-column-width=3.8mm]
  S[table-format=2.1, table-column-width=3.8mm]
}
\hline 
& \multicolumn{8}{c|}{\textbf{SSQ}} & \multicolumn{8}{c}{\textbf{IPQ}} \\ 
& \multicolumn{2}{c}{Total} & \multicolumn{2}{c}{Nausea} & \multicolumn{2}{c}{Oculomotor} & \multicolumn{2}{c|}{Disorientation} & \multicolumn{2}{c}{GP} & \multicolumn{2}{c}{SP} & \multicolumn{2}{c}{INV} & \multicolumn{2}{c}{ER} \\
& {M} & {SD} & {M} & {SD} & {M} & {SD} & {M} & {SD} & {M} & {SD} & {M} & {SD} & {M} & {SD} & {M} & {SD}\\
\hline 
\textbf{SilverCycling} &{7.5} & {9.0} & {6.0} & {10.4} & {3.8} & {8.3} & {12.2} & {10.0} & {5.9} & {1.1} & {5.2} & {0.7} & {4.9} & {0.5} & {3.9} & {0.9}  \\
\textbf{JC} &{9.8} & {9.4} & {8.3} & {10.4} & {4.3} & {6.8} & {16.5} & {10.4} & {5.7} & {1.1} & {4.8} & {0.7} & {4.6} & {0.7} & {3.6} & {0.9}\\ 
\hline 
\end{tabular}
\caption{Summary of results of SilverCycling and JC: The table presents mean and standard deviation for SSQ scores and its sub-scores, along with median values for the Igroup Presence Questionnaire (IPQ) scores by subscale. The subscales include General Presence (GP), Spatial Presence (SP), Involvement (INV), and Experienced Realism (ER). SSQ scores between 20 and 30 indicate minimal to moderate motion sickness, while scores above 40 suggest significant simulator discomfort. IPQ scores range from 1 (lowest) to 7 (highest) across all subscales.}
\label{tab:SSQ}
\end{table}

\subsection{Perceived Workload. }
The average NASA-TLX scores in all subscales showed no significant differences between the two locomotion and no correlation with task performance. Compared to JC, SilverCycling had lower mental ($M=3.25<3.81, p=0.12>0.05$), physical ($M=1.56<1.81, p=0.38>0.05$), and temporal ($M=1.19<1.38, p=0.18>0.05$) demand, required less effort ($M=3.25<3.81, p=0.44>0.05$) and led to better performance ($M=2.06<2.44, p=0.33>0.05$) and less frustration ($M=1.25<1.31, p=0.56>0.05$). 

Several participants (N=9) found JC unfamiliar, requiring heightened mental demand. With the SilverCycling system, a few participants (N=3) noted excessive mental demand on the locomotion tool during cycling. Additionally, some (N=2) reported high physical demand, suggesting its unsuitability for extended use.

\subsection{User Experience}
\textbf{Enjoyment, Safety and Comfort. }We outlined the statistical analysis of subjective feedback based on the 7-Likert scale and Wilcoxon Signed-Rank test results. The results showed that scores in \textit{Enjoyment} ($Z=-1.95$, $p=0.05$) and \textit{safety} ($Z=-2.06$, $p<0.05$) were significantly higher when navigating with SilverCycling compared to JC. The \textit{Enjoyment} attributed it to full body involvement, physical exercise, and the natural feel of cycling. For instance, P2 noted that \textit{``cycling is more fun as it involves the whole body, unlike JC which only uses fingers.''} A smaller group (N=2), expressed less enjoyment or even discomfort with JC P16 mentioned, \textit{``The controller is not very enjoyable as it feels tense and too fast.''} Participants (N=6) expressed the SilverCycling system as having higher perceived \textit{Safety}, citing its stability, ease of control, and lower risk of motion sickness. For instance, P8 mentioned, \textit{``I feel cycling's safer since it doesn't make you dizzy, and holding onto the handlebars makes you feel more secure. On the other hand, using JC can be a bit risky because of those sudden stops that can shoot up your blood pressure and make you feel dizzy.''} One participant (N=1), P4, reported feeling unsafe with both techniques, attributing this sense of insecurity to motion sickness experienced with each method. No significant difference in \textit{Comfort} when using SilverCycling locomotion as opposed to JC.

\textbf{Preference. }Participants (N=13) expressed a strong preference for using the SilverCycling system for navigation, which they praised for its intuitive interaction and physical engagement. In contrast, one participant (N=1) preferred JC, because it required less physical effort and provided a relaxed locomotion experience. Among the participants who preferred the SilverCycling system, nine particularly valued its physical engagement and benefits, in exercising. P2 stated, \textit{``I prefer the SilverCycling system for its bodily movement and exercise benefits.''}  Familiarity was another factor, with five participants favoring it for its intuitive interaction, contributing to a more comfortable experience. P3 reflected his sentiment: \textit{``Cycling feels more familiar to me, offering relaxation and a sense of balance.''} Two participants specifically mentioned its realism and immersion, as P12 noted: \textit{``I prefer cycling for its natural feel in turning.''}

The participants favoring JC appreciated its lower physical demands, precision, and novelty. As P1 described, \textit{``JC requires less physical effort, and I prefer the more relaxing approach.''} Participants who found merit in both systems noted their distinct advantages. P9 commented on JC's flexibility: \textit{``JC is appealing for its greater control flexibility.''} P11, embracing innovation and new technologies, said, \textit{``I'm enthusiastic about new techniques; JC feels more modern and technologically advanced.''} 

\section{DISCUSSION AND DESIGN CONSIDERATION}
In our study, we embarked on an initial exploration of natural motion-based VR locomotion techniques for older adults, addressing the RQ. First, we examined how natural motion-based, bike-based locomotion, impacts the spatial orientation among older adults. Second, we explored the design principles that could optimize these VR locomotion techniques to enhance the spatial orientation and overall user experience for older adults. 


\subsection{Potential Factors}
\subsubsection{Physical Motion Cues}
Our study indicated a significant enhancement in spatial orientation ability for older adults when using the SilverCycling system in the open-road urban environment. We attribute this enhancement primarily to the system's provision of physical motion cues akin to natural movement, a stark contrast to the limited body movement observed with JC's use. Notably, participants exhibited increased rotation angles in head movements and a higher frequency of lateral gaze when using SilverCycling compared to JC. These natural head movements in SilverCycling are likely to offer proprioceptive and vestibular cues, thereby enhancing spatial understanding and updating. This is in line with previous research, which highlights the positive effects of such motion cues on navigation performance \cite{bakker1999effects, bowman20043d, ruddle2006efficient, riecke2010we}. As Bowman et al. suggested, the physical translation produces vestibular self-motion information which furthers the walker’s spatial understanding \cite{bowman20043d}. Additionally, Riecke's research that considerable navigation improvements can already be gained by allowing for full-body rotations \cite{riecke2010we}. Conversely, the absence of these physical motion cues has been shown to impair spatial updating \cite{presson1994updating, wraga2004spatial}. According to participants' reports, the SilverCycling experience notably enhanced their ability to understand and navigate the virtual environment, a benefit largely attributed to the system's emphasis on physical engagement. A distinctive feature of this engagement is the requirement for participants to rotate their heads and shoulders in sync with the virtual rotation, providing a more immersive and effective spatial orientation experience.

\textbf{DC1: Incorporate Physical Movement Involving Body and Head Movement In Locomotion Techniques.} We propose the development of VR locomotion techniques that prioritize physical engagement for older adults, particularly focusing on techniques that involve rotation of the body and head. Our findings and existing literature suggest that body and head rotation techniques have the potential to enhance spatial orientation in VR environments \cite{bowman20043d, riecke2010we}. However, when implementing physical movement in VR locomotion for older adults, it is crucial to prioritize safety and moderate physical demands. 

\subsubsection{Familiarity}
Our results suggested that familiarity affects both older adults' spatial orientation and user experiences. Based on the demographic data concerning proficiency and frequency of cycling and JC, along with our interview findings, many participants showed greater familiarity and skill with cycling interactions. Conversely, they reported that joystick control required additional focus due to its unfamiliarity. This indicated that the high familiarity allowed them to focus more effectively on their orientation and the visual content presented in the virtual environment, instead of focusing on the locomotion techniques, hence improving the overall performance. These observations align with existing research, which suggests that familiarity can notably enhance the effectiveness, efficiency, and satisfaction of older adults when interacting with new technologies \cite{boger2013examining, leonardi2008designing}. Research indicates that younger generations, more familiar with digital technologies than previous cohorts, may shape future older adults' preferences for digital interfaces like JC used in gaming and other applications \cite{czaja2016designing}. As technology becomes more ubiquitous, Charness and Boot suggest that the learning curve for new interfaces may lessen, potentially making JC more accessible and appealing to these future older adults \cite{charness2009aging}. Therefore, although this study highlights the benefits of a bike-based locomotion interface, future research should re-evaluate the preferences and performance of older adults as their technological familiarity evolves.

\textbf{DC2: Implement Intuitive Motion Metaphor in the Design of VR Locomotion Techniques.} According to the findings, familiarity may be one of the factors contributing to enhanced spatial orientation and user experiences. Familiar movements such as walking, cycling, or rowing, along with intuitive metaphors like skiing or skating, can be translated into VR locomotion techniques \cite{Wei2023dragon}. These well-known physical activities may allow users to navigate and interact in the virtual space using motions that are already ingrained in their muscle memory. By leveraging these familiar and intuitive motion metaphors, VR experiences can become more accessible and enjoyable for all users, particularly older adults who may be less familiar with digital and VR technologies. However, leveraging familiar movements can enhance spatial orientation and user experiences; these activities might require physical demand, which is also a frequently mentioned drawback of SilverCyling by the participants in the interview. Additionally, as demonstrated in our pilot study, simulating walking with unnatural methods resulted in significant learning difficulties and physical burdens for older adults. Therefore, designing VR locomotion techniques requires a delicate balance between reducing physical effort and preserving effective interaction, ensuring that the metaphor does not inadvertently increase the older adults' learning curve.  

\subsubsection{Motion Sickness}
Our results suggested that VR motion sickness influences spatial and visual information perception and other experiences. While our study did not reveal statistically significant differences in VR sickness as measured by SSQ, we observed a trend suggesting lower levels of VR sickness among participants who used the SilverCycling system. This issue of motion sickness was notably more pronounced in JC usage. Several participants reported increased dizziness, which they attributed to the challenges in managing the turning and abrupt stopping actions when using JC. Consequently, several participants reported lower scores in subjective safety and comfort, attributing these scores to feelings of dizziness. 

Additionally, we observed a positive correlation between the severity of motion sickness induced by SilverCycling and improved spatial orientation. This relationship seems to stem from the specific nature of motion sickness triggered by SilverCycling, which is intimately associated with heightened head movements. We found that using SilverCycling promoted head rotation and lateral gazing during turns. Such movements are likely to contribute to improved spatial orientation by providing additional vestibular and visual information \cite{howard1966human}. However, it is important to note that this increase in head motion may also amplify the severity of motion sickness \cite{gordon1996vestibulo}. 

Conversely, our study found that participants who experienced stronger motion sickness during interacting with JC did not show any improvement in spatial orientation. This phenomenon may be linked to the specific factors causing motion sickness in these scenarios. Unlike SilverCycling, where participants engage in continuous turning with significant head and body movements, JC's usage typically involve non-continuous turning with minimal physical movement, failing to provide consistent vestibular cues. The primary cause of motion sickness in these instances is abrupt starts or stops during non-continuous turning. Moreover, participants using JC often exhibited a unidirectional gaze. Consequently, the type of motion sickness induced by these unexpected movements was not associated with an enhancement in spatial orientation.

The interesting findings could offer new insights into the relationship between motion sickness and spatial orientation, and guide the design of more effective VR interactions. Further exploration of the underlying mechanisms, including physical movement, visual input, and the vestibular sense of balance, is needed to identify mediating factors that could alleviate motion sickness and improve spatial orientation.

\textbf{DC3: Adopt Soft-start and Soft-stop Mechanisms to Reduce Motion Sickness.} Our research showed that abrupt movements in VR, such as quickly pulling the joystick to the farthest extent and then stopping, would cause dizziness and discomfort in older users, potentially raising their blood pressure. Moreover, motion sickness induced by this factor did not correlate with the improvement of spatial orientation in our study. Therefore, We suggest reducing sudden movements in VR locomotion and implementing smoother, gradual start and stop mechanisms. For instance, when using the SilverCycling system, participants can gently adjust their pace at the beginning and the end. This method can help lower motion sickness risks, enhancing comfort and safety in VR for older users. 

\textbf{DC4: Separate Different Controls to Reduce Motion Sickness.} Our study revealed that participants often triggered incorrect functions by mistake when using JC, leading to motion sickness. For example, some inadvertently turned left when intending to turn right. For instance, some participants would inadvertently turn left when intending to turn right. This problem was especially prominent with JC, where functions were condensed into a small joystick. This design is not user-friendly, especially for older adults unfamiliar with such devices. Additionally, turning and moving forward simultaneously were reported by some participants as causing dizziness. According to Kuiper et al.'s findings, unpredictable motion causes more motion sickness \cite{kuiper2020knowing}. To avoid this, we recommend separating essential functions such as moving forward, turning left, and turning right into distinct and dedicated buttons or interactions. 

\section{LIMITATION AND FUTURE WORK}
\subsection{Enhancing Interaction in Natural Motion-Based Locomotion Systems for VR Environments}
The SilverCycling system enhances spatial orientation for older adults in the urban VR environment more effectively than JC. However, it limits interactions due to occupied hands, impacting the overall VR experience. While it supports basic navigation, it hinders complex interactions required for fully exploring VR, such as object manipulation. To address this, integrating additional interfaces like buttons could allow users to engage with the VR environment without disrupting locomotion \cite{zeleznik2002pop}. Furthermore, incorporating multimodal inputs such as gesture, eye gaze, and voice can make the system more intuitive and accessible \cite{piumsomboon2017exploring}. Selecting appropriate interaction modalities should be informed by a comprehensive analysis of task requirements and user preferences to ensure an optimal blend of interaction mechanisms and locomotion strategies.

\subsection{Assessing the Impact of Long-Term Use on Natural Motion-Based Locomotion Systems}
The use of natural motion-based locomotion systems like the SilverCycling system places considerable physical demands on users, which can pose substantial challenges during extended periods of use. This is especially true for older adults and individuals with limited mobility or physical strength. Over time, the requirement for continuous physical movement inherent in these systems can lead to increased fatigue and discomfort, thereby diminishing the quality of the VR experience \cite{10.1145/2967934.2968105}. Given these potential concerns, it is essential to rigorously assess the impact of long-term engagement with the SilverCycling system to determine its effectiveness and to ensure a positive user experience.

\subsection{Expanding Environmental Contexts in Locomotion Technique Evaluation}
The study's evaluation was confined to a single setting, omitting the examination of the locomotion technique across diverse environments such as indoor spaces, pedestrian thoroughfares, and public areas. These settings, which often present more complex interaction dynamics than the open-road environment used in the study, could lead to different user behaviors and experiences. For example, navigating public spaces may require users to avoid obstacles, necessitating alternative interaction strategies. The exclusion of these varied environments limits the depth of the locomotion technique's assessment. Future studies should explore a broader range of environments and interactive challenges to gain a more comprehensive understanding of the technique’s efficacy in diverse scenarios, especially for older adults.

\subsection{Expanding Demographic Constraints of the Future Study in Locomotion Technique}
Our study's focus on older adults with mobility capabilities and without severe visual impairments may limit the generalizability of our findings to broader demographics. While the sample size and methodology were appropriate for our objectives, they might not capture the full diversity of experiences and preferences within the wider VR user base. Future studies should include a more varied participant pool to ensure results are representative of all VR users.

\subsection{Expanding VR Locomotion Design for Older Adults}
The VR setups used in our study might not represent the wide range of VR technologies available, more types of locomotion techniques should be tested in the future. Moreover, our pilot study findings indicate that older adults face difficulties when using the current ODW and WIP methods. Consequently, there is a need to design and examine walking-based locomotion techniques that are tailored to the capabilities and needs of the older adult demographic.

\section{CONCLUSION}
In conclusion, our study initiated an exploration into natural motion-based locomotion techniques for older adults enhancing their spatial orientation performance in the open-road urban environment and overall user experience in VR, particularly focusing on bike-based locomotion. To address this, we first developed the SilverCycling system, a locomotion system that provides consistent turning and forward motion optimized for HMDs. The system was designed for ease of navigation in virtual cities among older adults. The results indicate that the SilverCycling system may enhance older adults' spatial orientation in the open-road urban environment when compared to JC. Moreover, the acceptance and usability of the SilverCycling system among the participants highlight its potential as a beneficial tool in VR navigation in the open-road urban environment. The study's results also provide insights into optimizing VR locomotion techniques to improve the spatial orientation and the overall user experience for this demographic.

\bibliographystyle{ACM-Reference-Format}
\bibliography{main.bib}

\end{document}